\documentclass[preprint,journal]{vgtc}       
\ieeedoi{10.1109/TVCG.2019.2934668}

\ifpdf
  \pdfoutput=1\relax                   
  \usepackage{graphicx}                
  \DeclareGraphicsExtensions{.pdf,.png,.jpg,.jpeg} 
\else
  \usepackage{graphicx}                
  \DeclareGraphicsExtensions{.eps}     
\fi%

\graphicspath{{images/}} 

\usepackage{microtype}                 
\PassOptionsToPackage{warn}{textcomp}  
\usepackage{textcomp}                  
\usepackage{mathptmx}                  
\usepackage{times}                     
\usepackage{cite}                      
\usepackage{tabu}                      
\usepackage{booktabs}                  

\usepackage{graphicx}
\usepackage{times}
\usepackage{cite}
\usepackage{listings}
\usepackage{amsmath}
\usepackage{amssymb}
\usepackage{subfigure}
\usepackage[draft]{pgf}
\usepackage{xspace}
\usepackage{lipsum}
\usepackage{enumerate}
\usepackage{mathtools}
\usepackage{color}
\usepackage{array}
\usepackage{tabularx}
\usepackage{makecell}

\definecolor{myred}{RGB}{190,30,45}
\definecolor{myblue}{RGB}{46,49,146}
\definecolor{myyellow}{RGB}{250,167,26}

\newcommand{\ie}{\emph{i.e.}\xspace}
\newcommand{\eg}{\emph{e.g.}\xspace}

\newcommand{\myparagraph}[1]{\noindent \textbf{#1}}
\newcommand{\denselist}{\itemsep 0pt\parsep 1pt\partopsep 0pt}

\newcommand{\visflow}{VisFlow\xspace}
\newcommand{\flowsense}{FlowSense\xspace}
\newcommand{\sempre}{SEMPRE\xspace}
\newcommand{\query}[1]{``{\it{#1}}''}
\newcommand{\var}[1]{\langle{#1}\rangle}
\newcommand{\parwidth}[1]{\makecell*[{{p{\linewidth}}}]{#1}}
\newcommand{\footlabel}[2]{%
    \addtocounter{footnote}{1}%
    \footnotetext[\thefootnote]{%
        \addtocounter{footnote}{-1}%
        \refstepcounter{footnote}\label{#1}%
        #2%
    }%
    \hspace{0pt}$^{\ref{#1}}$%
}
\newcommand{\footref}[1]{\hspace{0pt}$^{\ref{#1}}$}
\newcommand\nodot[1]{}

\onlineid{1043}

\vgtccategory{Research}
\vgtcpapertype{system}

\title{FlowSense: A Natural Language Interface for Visual Data Exploration within a Dataflow System}

\author{Bowen Yu and Cl\'{a}udio T. Silva \textit{Fellow, IEEE}}
\authorfooter{
\item
 Bowen Yu is with New York University. E-mail: bowen.yu@nyu.edu.
\item
 Cl\'{a}udio T. Silva is with New York University. E-mail: csilva@nyu.edu.
}

\shortauthortitle{Biv \MakeLowercase{\textit{et al.}}: Global Illumination for Fun and Profit}

\abstract{
Dataflow visualization systems enable flexible visual data exploration by allowing the user to construct a dataflow diagram that composes query and visualization modules to specify system functionality.
However learning dataflow diagram usage presents overhead that often discourages the user.
In this work we design FlowSense, a natural language interface for dataflow visualization systems that utilizes state-of-the-art natural language processing techniques to assist dataflow diagram construction.
FlowSense employs a semantic parser with special utterance tagging and special utterance placeholders to generalize to different datasets and dataflow diagrams.
It explicitly presents recognized dataset and diagram special utterances to the user for dataflow context awareness.
With FlowSense the user can expand and adjust dataflow diagrams more conveniently via plain English.
We apply FlowSense to the VisFlow subset-flow visualization system to enhance its usability.
We evaluate FlowSense by one case study with domain experts on a real-world data analysis problem and a formal user study.
} 


\CCScatlist{ 
 \CCScat{H.5.2}{Information Interfaces and Presentation}%
{User Interfaces}{Graphical user interfaces (GUI)}{Natural language};
}

\teaser{
\includegraphics[width=.9\linewidth]{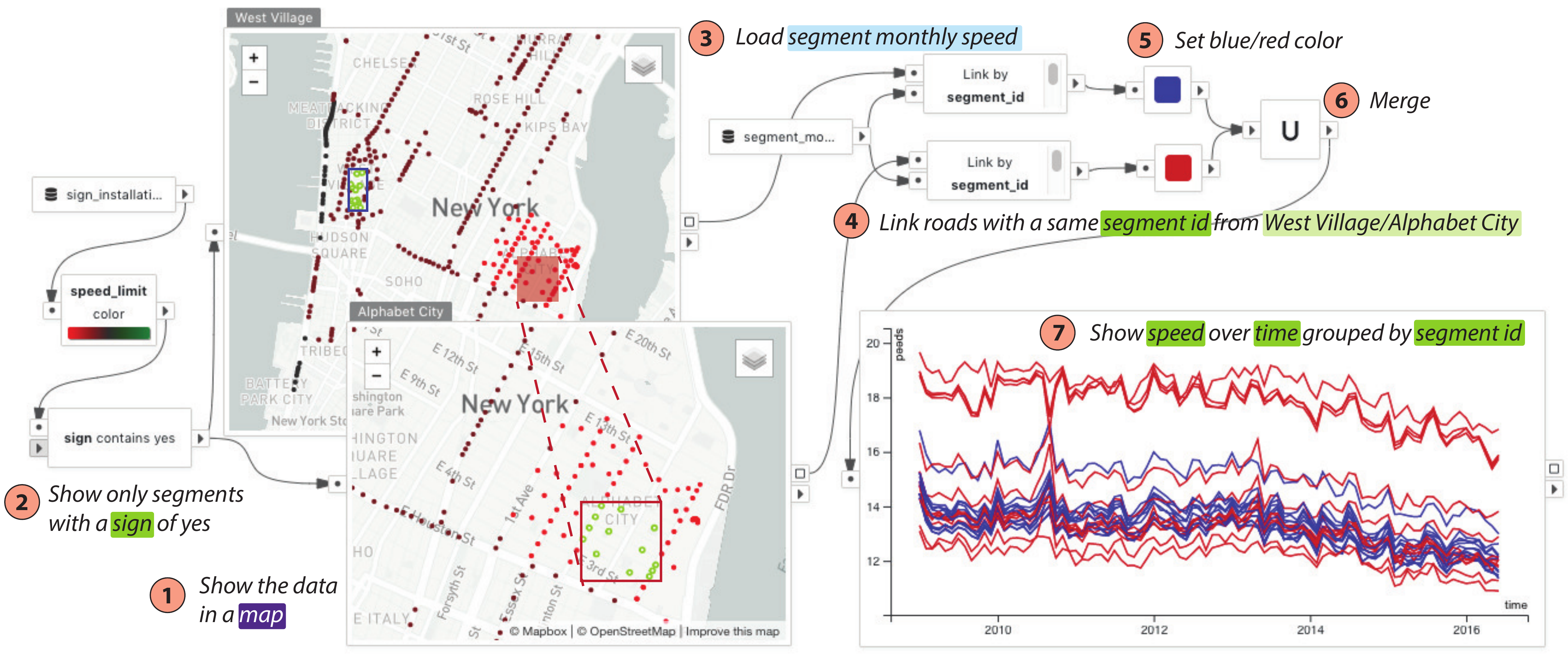}
\centering
\caption{
Using \flowsense for a comparative study on the street speed changes between two slow zones: West Village (blue) and Alphabet City (red).
The analysts start by drawing locations of speed limit signs, which appear as dots with speed limits encoded by color.
The selected speed limit signs are interactively linked with the line chart that shows the changes of average vehicle speed over time on the corresponding streets.
All diagram elements are created via \flowsense.
The NL queries shown are executed in the numbered order.
\flowsense processes rich dataflow context and allows the user to reference dataflow elements at different specificity levels, \eg node labels, node types, or implicitly.
}
\label{fig:visionzero-b}
}

\vgtcinsertpkg

\begin{document}



\firstsection{Introduction}
\label{sec:intro}

\maketitle

Natural language interfaces (NLI) for data visualizations seek better usability of visualization solutions by introducing natural language (NL) query support.
Compared with visualization systems that only support traditional mouse/keyboard interactions,
systems with NLI require less prior knowledge about their functionality and usage details to work with.
Latest research has progressed in visualization-oriented NLIs~\cite{TongG15,SetlurV16,HoqueE18}.
Most of these interfaces present a single visualization answer that can be interacted with (possibly with a few auxiliary views and widgets).
The user does not have the opportunity to specify the relationships between multiple visualizations.
However, practical data analysis tools often have multi-view linked visualizations, for which the design of an NLI becomes more challenging.

Dataflow visualization systems (DFVS) have been proposed to achieve larger analytical flexibility~\cite{UpsonC89, ParkerS95, BavoilL05}.
These general-purpose visualization toolkits allow the user to draw a dataflow diagram that composes system modules to process and visualize data.
It has been shown that DFVS can help build data analysis environments with multi-view linked visualizations that adapt to different domains~\cite{WaqasJ13,YuB17}.
Despite the flexibility, a DFVS often has higher learning overhead, due to its dataflow complexity, than a bespoke visualization application in which system components have pre-defined connections.
The user must be proficient with the underlying DFVS modules to effectively use it.

In this work we propose \flowsense, a novel NLI that seeks to benefit both from the usability of NL and the analytical flexibility of DFVS.
\flowsense uses semantic parsing to support NL queries that manipulate multi-view visualizations produced by a dataflow diagram.
The NL capability may help reduce the overhead of learning dataflow and simplify the interactions of dataflow diagram construction.
The \flowsense input box utilizes special utterances tagged by the underlying parsing algorithm to provide real-time feedback on what the system sees and understands.
To demonstrate the application of \flowsense, we build it on top of the recent dataflow system \visflow created by Yu et al.~\cite{YuB17}.
We choose \visflow because it focuses on generating linked visualizations that have good interactivity and support brushing and linking,
which are two essential aspects of visual data exploration.
With the integration of \flowsense, dataflow diagram editing becomes more intuitive in \visflow, and consequently the user can use the DFVS more efficiently.
The contributions of this work are summarized as follows:

\vspace{-1.5mm}
\begin{enumerate}[1)] \denselist
\item 
We propose \flowsense, a novel NLI for visual data exploration within a DFVS.
\flowsense uses NL to reduce the dataflow learning overhead and improve the DFVS usability, while taking advantage of the flexibility of a DFVS.

\item 
We exemplify a generalizable approach of applying state-of-the-art semantic parsing techniques to create a grammar that is tailored for a DFVS.
In particular, \flowsense employs special utterance tagging and special utterance placeholders to be aware of the dataflow context,
and make its grammar independent of datasets, dataflow diagram elements, and analytical tasks.
The identified special utterances are presented interactively as the user types the query.
Such a design echoes the underlying parsing state to the user.
It not only helps the user understand the query semantics behind the scene, but also is useful for identifying errors and resolving ambiguity.

\item 
We demonstrate that \flowsense is able to support NL queries for the majority of dataflow diagram editing operations in \visflow.
We showcase the application of \flowsense by a case study with domain experts on studying the traffic speed reduction based on NYC taxi trip data.
We further conduct a formal user study to evaluate the proposed NLI.
We measure the task completion time, collect user feedback, and analyze the NL query logs to identify the strengths and weaknesses of \flowsense.
\end{enumerate}
\vspace{-1.5mm}
Details on the \flowsense grammar and its implementation can be found in the appendix and the \flowsense GitHub repository\footnote{https://github.com/yubowenok/flowsense}.
\vspace{3mm}


\section{Related Work}
\label{sec:related}

\subsection{Dataflow Visualization System (DFVS)}
Dataflow systems enable the user to configure system functionality by drawing a dataflow diagram that defines how the system modules interact with each other.
While dataflow systems are effective in fields other than data visualization such as computational workflow design~\cite{KatherineW13,SPSSModeler,Knime},
we focus on dataflow systems for visualization purposes in this section.
Previous DFVS have demonstrated the effectiveness of using dataflow to render scientific data~\cite{UpsonC89,ParkerS95,HaeberliPE88}
and manage volume rendering pipelines~\cite{BavoilL05,MeyerSpradowJ09}.
Dataflow systems that pass only data subsets (versus program method arguments) yield simpler dataflow diagrams and lower learning overhead~\cite{RobertsJC98, RobertsJC99}.
ExPlates~\cite{WaqasJ13} and \visflow~\cite{YuB17} present embedded visualizations in their dataflow, and focus on interactive information visualization.
Most dataflow systems support diagram editing in a drag-and-drop manner.
However, it is observed that even with drag-and-drop interfaces, users may often have difficulty in translating their intention to system operations~\cite{GrammelL10}.
In this work we design \flowsense to further simplify dataflow diagram construction, so that the user can intuitively use dataflow and make the most of the analytical capability of a DFVS.
In particular, we build \flowsense for \visflow, as its subset flow model supports many of the low-level visual data analysis tasks~\cite{ShneidermanB96, AmarR05}, such as characterizing distribution, finding extremum, etc. %

\subsection{NLI for Data Visualization}
Extensive research has been devoted to NLIs for decades.
These interfaces address NL queries that otherwise have to be manually translated to formal query languages, \eg SQL.
A few examples are the interfaces for querying XML~\cite{YLi07}, entity-relational database~\cite{AndroutsopoulosI95, YinP16}, and sequence translator to SQL~\cite{ZhongV17}.
NLIs for data visualizations answer the queries by presenting visual data representations.
Compared with other interfaces that simply return a numerical answer or a set of database entries,
visualization NLIs present results that are more human-readable. 
Cox et al.~\cite{CoxK01} design the Sisl service within the InfoStill data analysis framework.
The service asks a series of NL questions to complete an unambiguous query.
The Articulate system~\cite{SunY10} uses a Graph Reasoner to select proper visualizations to answer a query.
DataTone~\cite{TongG15} addresses query ambiguity by showing ambiguity widgets along with the main visualization so that the user is able to switch to desired alternative views.
Eviza~\cite{SetlurV16} and Evizeon~\cite{HoqueE18} further improve the user experience by allowing for conversation-like follow-up questions.
Fast et al.~\cite{FastE17} propose a conversational user interface called Iris that may perform analytical tasks and plot data upon requests in dialogues.
Kumar et al.~\cite{KumarA16} also propose a dialogue system for visualization.
Orko~\cite{SrinivasanA18} is an NLI designed for visual exploration of network data.
Dhamdhere et al.~\cite{DhamdhereK17} design Analyza that provides database-based NL query and visualizations.
Srinivasan et al.~\cite{SrinivasanA17} provide a summary and comparison of the majority of these NLIs.
Several commercial tools integrate NLIs.
IBM Watson Analytics~\cite{WatsonAnalytics} and Microsoft Power BI~\cite{PowerBI} provide a list of relevant data and visualizations to an NL question, from which the user may choose to continue an analysis.
Wolfram Alpha~\cite{WolframAlpha} supports knowledge-based Q\&A and is able to plot the results.
ThoughtSpot~\cite{ThoughtSpot} enables interactive search in a relational database, and provides multiple types of visualizations for the database.
The NLI design for data visualization has two challenges:
First, modern natural language processing (NLP) techniques cannot yet understand well arbitrary NL input due to the complex nature of NL.
User queries are apt to be free-form and ambiguous;
Second, choosing a proper visualization to answer an analytical question is non-trivial as there can be multiple possible visual representations~\cite{MackinlayJ07}.

\subsection{Comparison with Other NLIs}

\noindent\flowsense makes a distinction from the other interfaces as it is to our best knowledge the first NLI to address a dataflow context.
We set the scope of \flowsense to focus on assisting dataflow diagram construction, rather than to directly answer free-form analytical questions or seek a best visualization for a given query.
We believe such an approach is beneficial in several aspects:\\
\myparagraph{Capability:} The analytical capability of \flowsense is rooted in the design of the DFVS.
The outcome of \flowsense is a complete, interactive, and iterative visual data exploration process supported by the DFVS, rather than a single visualization that only answers one particular query as in many other interfaces.
Dataflow also naturally preserves analysis provenance~\cite{FreireJ06},
allowing the user to frequently revisit and reassess the current workflow.
The diagram created by \flowsense explicitly keeps the user's preference and intention from previous queries,
which must otherwise be maintained by a model behind the scene~\cite{TongG15,SetlurV16}.

\myparagraph{Usability:} 
\flowsense integrates real-time presentation of tagged special utterances in the interface that reflect the state of the underlying semantic parser and help the user understand the dataset and dataflow present in the system (\autoref{sec:dataflow-context}).
This is a novel design that facilitates the user's understanding of the NLI behavior, as in most other NLIs the parsing feedback is only given after the query is submitted.
The auto-completion suggestions of \flowsense also present special utterance tags so that the user may better understand the expected query components.
Consequently, \flowsense may ease DFVS usage and make DFVS more accessible.
Our case study and user study (\autoref{sec:evaluation}) show that \flowsense improves the DFVS usability, and its convenience is desirable by both novice and experienced \visflow users.
Besides, the DFVS is able to recover from errors more easily as the user always has full control over the system.
However in other interfaces the user has to mostly rely on the behavior of the NLI and can hardly make corrections in case of misinterpretation.

\myparagraph{Feasibility:} 
The scope of assisting dataflow diagram construction is well defined and practicable.
Even state-of-the-art NLP techniques have limited success in understanding an arbitrary query.
Because each query is expected to update dataflow diagram and the user decides what the system should do and what visual representation to apply,
\flowsense can produce more expected results and give better user experience under a well-defined scope.
The mixed-initiative design mitigates the ambiguity problem.
The DFVS users in our case study and user study are all able to understand the scope of \flowsense and use \flowsense effectively.


\begin{table*}[t]
\centering
\begin{tabular*}{\linewidth}{cm{.1\linewidth}p{0.2\linewidth}p{0.2\linewidth}p{180px}}
\toprule
{\bfseries \#} & {\bfseries \makecell{Function}} & {\bfseries Sample Queries} & {\bfseries Description} & {\bfseries Sample Sub-Diagram}\\
\midrule
A & \makecell{Visualizing} &
\parwidth{\itshape{Show a scatterplot of mpg and horsepower}} &
\parwidth{Present the data in a visualization} &
\makecell{\includegraphics[width=\linewidth]{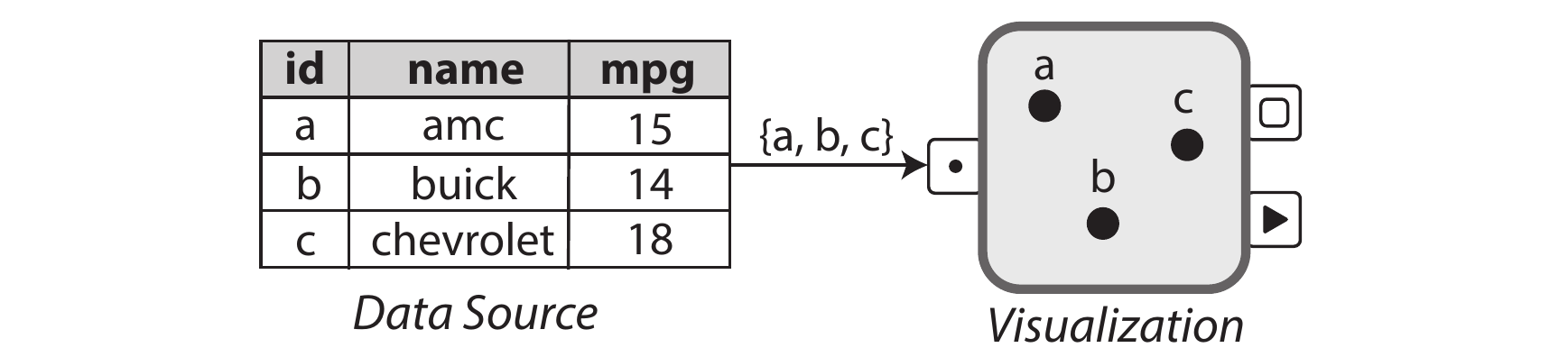}}
\\\hline

B & \makecell{Visual\\Encoding} &
\parwidth{\itshape{Encode mpg by red green color scale}} &
\parwidth{Map data attributes to visual channels} &
\makecell{\includegraphics[width=\linewidth]{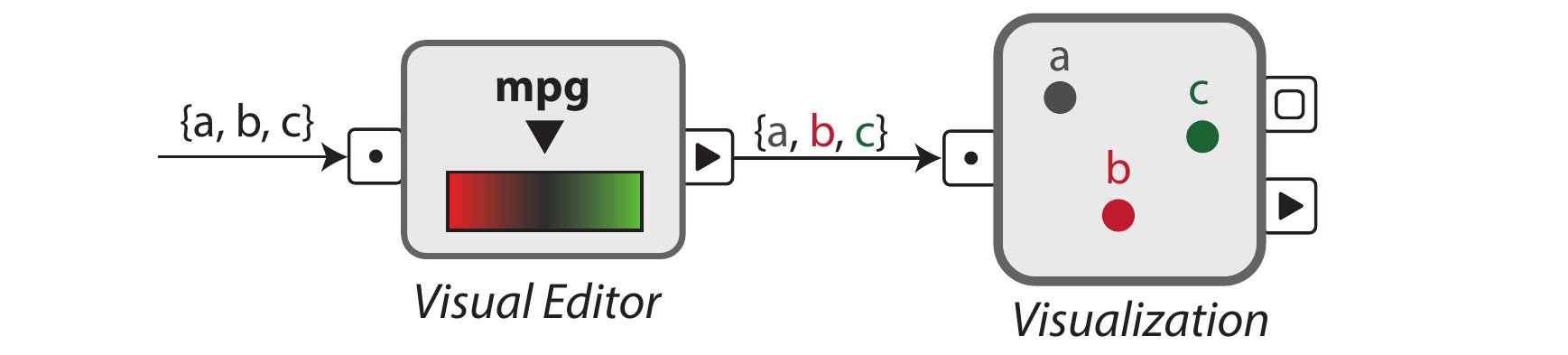}}
\\\hline

C & \makecell{Filtering and\\Finding\\Extremum} &
\parwidth{\itshape{Find all cars with mpg between 15 and 20};\\ \itshape{List five cars with maximum mpg}} &
\parwidth{Filter data items and locate extremums and outliers} &
\makecell{\includegraphics[width=\linewidth]{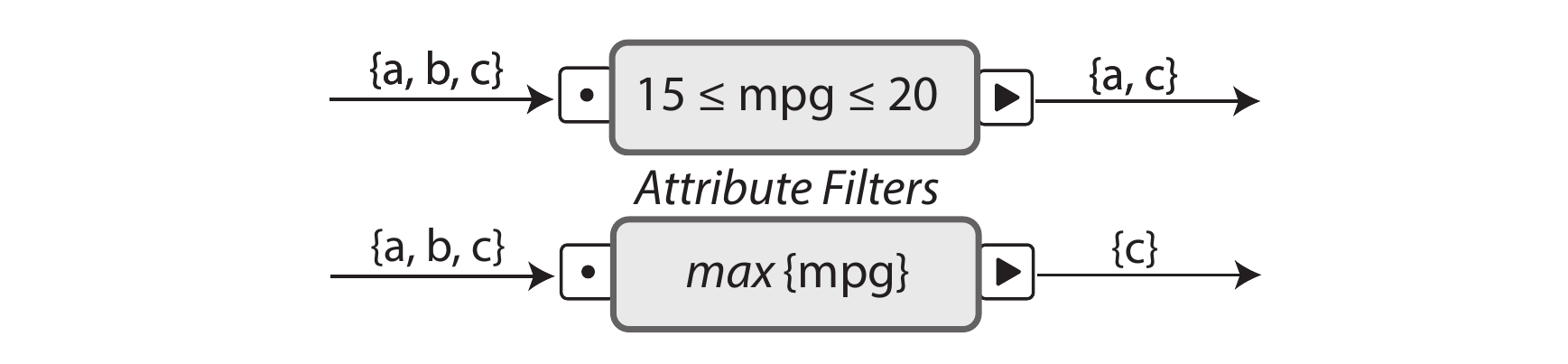}}
\\\hline

D & \makecell{Subset\\Manipulation} &
\parwidth{\itshape{Merge the cars with those from the scatterplot}} &
\parwidth{Refine and identify interesting subsets} &
\makecell{\includegraphics[width=\linewidth]{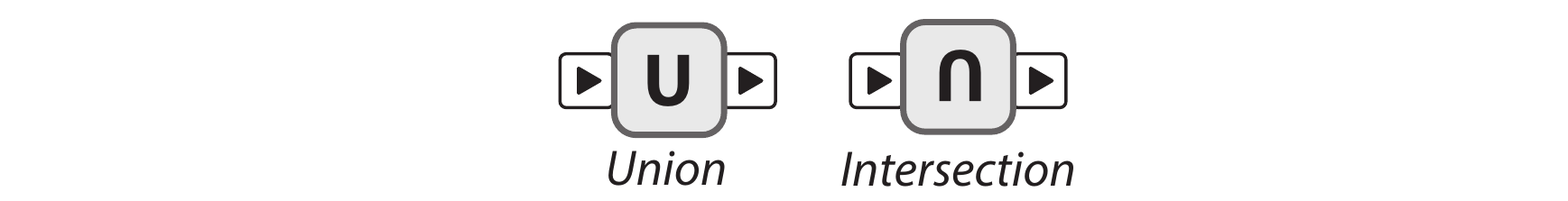}}
\\\hline

E & \makecell{Highlighting} &
\parwidth{\itshape{Highlight the selected cars in a parallel coordinates plot}}&
\parwidth{View the characteristics of one subset among its superset or another subset} &
\makecell{\includegraphics[width=\linewidth]{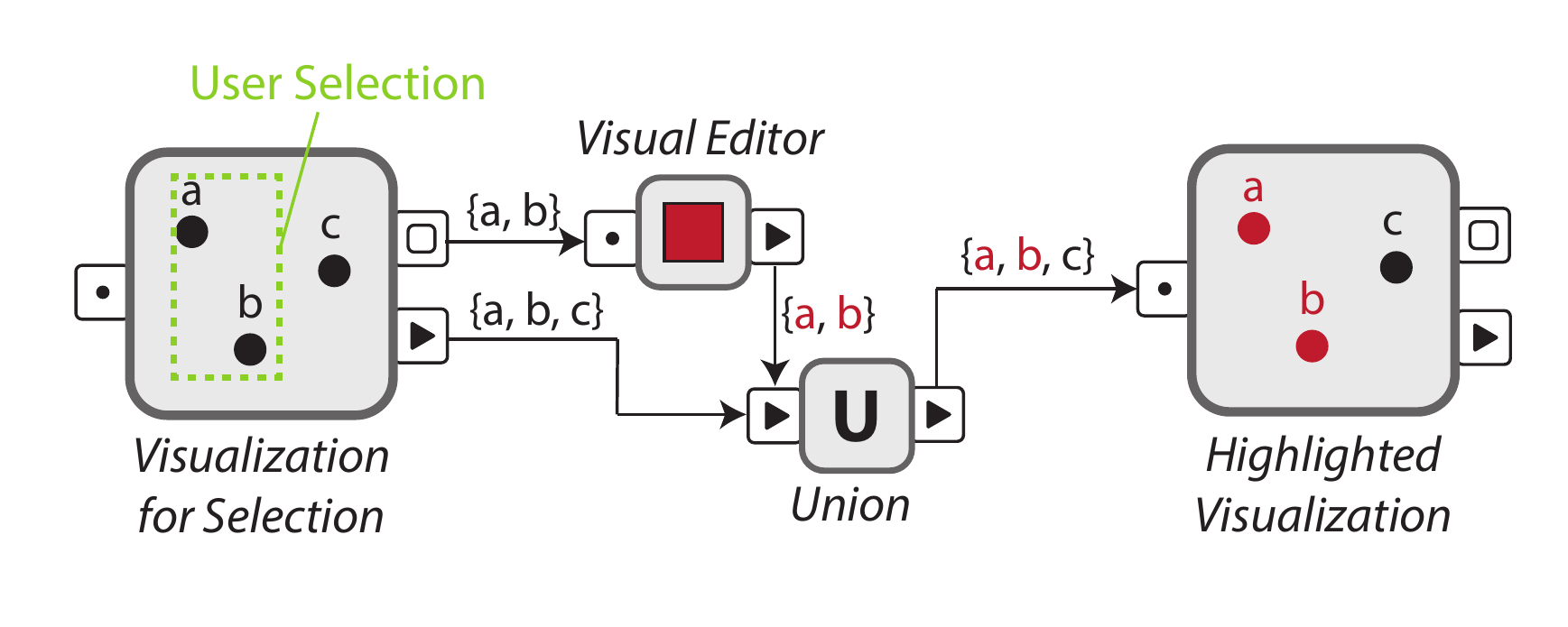}}
\\\hline

F & \makecell{Linking} &
\parwidth{\itshape{Link the cars with a same name from the sales table}} & 
\parwidth{Extract primary keys from one table and find their corresponding rows from another (heterogeneous) table} &
\makecell{\includegraphics[width=\linewidth]{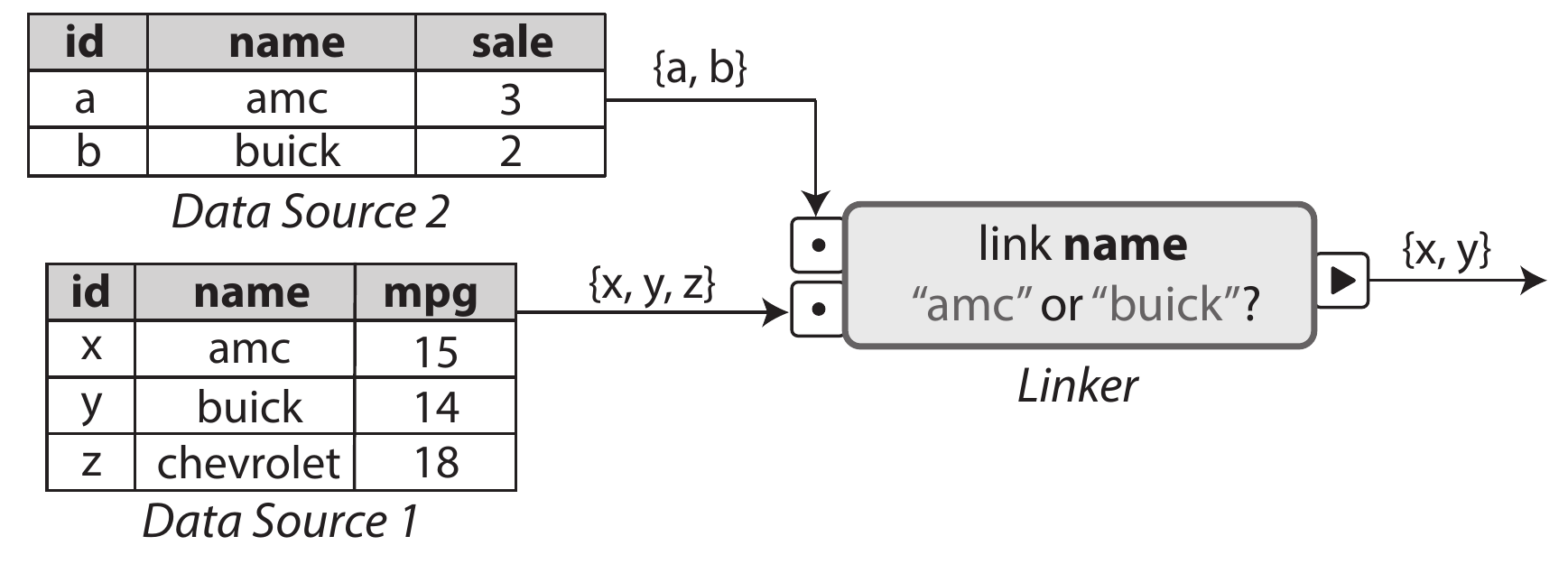}}
\\
\bottomrule
\vspace{-1mm}
\end{tabular*}
\caption{Six major categories of \visflow functions.
 These sub-diagrams are frequently used to compose more sophisticated diagrams that address analytical tasks.
 \flowsense aims at mapping NL input to one of these functions.
The illustration only shows one possible sub-diagram from each category and does not exhaustively list all the possible sub-diagram variations of the function options.
In practice the user can specify the function options via NL, \eg visualization type, filter type, etc.
Combinations of functions may apply.
}
\vspace{-2mm}
\label{tab:functions}
\end{table*}

\begin{figure*}[t]
\begin{center}
  \includegraphics[width=.9\textwidth]{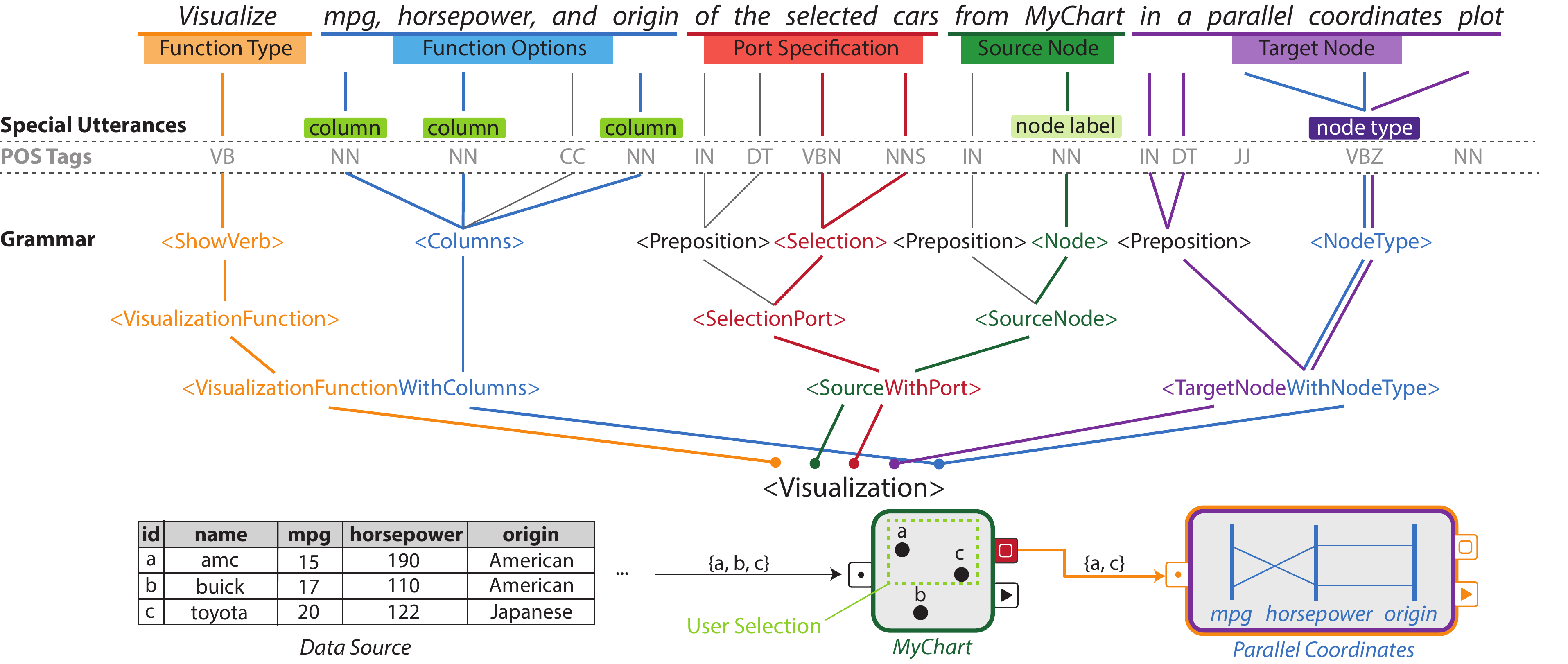} 
\end{center}
\vspace{-3mm}
 \caption{An example \flowsense query and its execution over the Auto MPG dataset.
The derivation of the query is shown as a parse tree in the middle.
The sub-diagram expanded by the query is illustrated at the bottom.
The five major components of a query pattern are underscored.
Each component and its relevant parts in the parse tree and the dataflow diagram are highlighted by a unique color.
The result of executing this query is to create a parallel coordinates plot of columns mpg, horsepower, and origin, with its input coming from the selection port of the node labelled MyChart.
 }
 \vspace{-2mm}
\label{fig:pattern}
\end{figure*}

\subsection{Semantic Parsing}
\flowsense uses semantic parsing to process NL input and map user queries to \visflow functions (\autoref{sec:functions}).
It depends on a pre-defined grammar that captures NL input patterns.
A semantic parser recursively expands the variables in the grammar to match the input query and can interpret the input based on the rules applied and the order of their application~\cite{BerantJ13}.
At a high level, the mapping performed by \flowsense can also be considered a classification task and addressed by classification algorithms~\cite{AllahyariM17}.
However we prefer semantic parsing because most classification approaches are supervised algorithms that require a large corpus of labeled examples.
Such training data are not available for DFVS.
Besides, compared with deep learning methods~\cite{LiD14,GoodfellowI16}, semantic parsing does not require heavy computational resources.

The \flowsense semantic parser is implemented within the Stanford \sempre framework~\cite{PasupatP15} and CoreNLP toolkit~\cite{ManningCD14}.
The CoreNLP toolkit integrates a comprehensive set of NLP tools including the Part-of-Speech (POS) tagger, Name-Entity-Recognizer (NER), etc.
A POS tagger identifies roles of words in a sentence, e.g. verb, preposition, adverb.
The \sempre framework employs a modular design in which different types of parsers and logical forms can be easily plugged-in.
The framework can quickly be adapted for domain-specific parser design~\cite{WangY15}.
We apply \sempre together with CoreNLP to the DFVS domain.
In particular, the \flowsense parser utilizes the POS tags produced by CoreNLP for processing special utterances and grammar matching.
The \flowsense grammar expects words with certain POS tags to appear in query parts.


\section{Semantic Parser}
\label{sec:design}
In this section, we define the building blocks of the semantic parser:
the \visflow functions that can be specified by NL,
the definition of the parsing grammar,
and the general query pattern the parsing algorithm expects.
For concept illustration we use the Auto MPG dataset\footnote{http://archive.ics.uci.edu/ml/datasets/Auto+MPG} throughout the paper,
which has information about cars in 9 columns, including mpg, horsepower, origin, etc.


\subsection{\visflow Functions}
\label{sec:functions}
To create an NLI for \visflow, we first studied a sample diagram set that includes 60 dataflow diagrams created by 16 \visflow users from their recorded \visflow sessions.
These diagrams cover a wide range of \visflow usage scenarios and deal with various types of datasets.
We identify a set of frequently appearing sub-diagrams and categorize them into six major categories as listed in \autoref{tab:functions}. 
The construction of these sub-diagrams are defined as the {\it\visflow functions}.
By implementing the \visflow functions, \flowsense essentially supports the building blocks of visual data exploration in \visflow so that analyses rendered by \visflow native interactions can be carried out with \flowsense.
These functions also reflect the fundamental analytical activity defined in information visualization task taxonomies~\cite{ShneidermanB96,AmarR05}.
\autoref{tab:functions} explains the usage of each \visflow function and shows several sample queries.

In addition to the six major categories, \flowsense also supports many utility functions such as adding/removing dataflow nodes/edges, undo/redo, loading datasets, etc.
Though these functions also enhance the usability of the system, we omit them here as they are indirectly related to visual data analysis.


\subsection{Dataflow Context and Special Utterances}
\label{sec:dataflow-context}
It is important to make the semantic parser aware of the dataflow context, such as the dataset loaded and the nodes in the dataflow diagram.
\flowsense extracts a special group of tokens called the {\it special utterances} from NL input.
Special utterances are words that refer to entities in the dataset or the dataflow diagram.
They are the arguments and operands of \visflow functions.
\flowsense recognizes table column names, node labels, node types, and dataset names as special utterances.
For the query shown in \autoref{fig:pattern}, \flowsense identifies ``mpg'', ``horsepower'', and ``origin'' as table columns, ``MyChart'' as a node label, and ``parallel coordinates'' as a node type.
The special utterances identified by \flowsense are shown in colored tags in the \flowsense input box (\autoref{fig:flowsense-ui}).
Each distinct color represents one special utterance type: green for table column, light green for node label, purple for node type, and light blue for dataset name.
The colors are applied consistently throughout the user interface.


\subsection{Grammar}
\flowsense applies a semantic parser to map an NL query to one of the \visflow functions based on an elaborate grammar designed for these functions.
The grammar is context-free~\cite{SipserM12} and formally defined as a 4-tuple $G = (V, \Sigma, R, S)$.
$V$ is a finite set of variables.
$\Sigma$ is a finite set of terminals.
A terminal represents an English word or phrase.
$R$ is the rule set that defines how a single variable matches an ordered list of terminals and variables (possibly itself in a recursive rule).
Below is an example rule:
\begin{align}\nonumber
\var{Visualization} \rightarrow \var{ShowVerb}~\var{Columns}~in~\var{VisualizationType}
\end{align}
In this rule, $\var{Visualization}$ is a high-level variable that matches a query that requests a visualization.
$\var{ShowVerb}$ matches a verb that has a meaning similar to ``show''.
$\var{Columns}$ matches one or more columns from the data.
$\var{VisualizationType}$ stands for a phrase that describes a visualization metaphor such as scatterplot or parallel coordinates.
The token ``in'' is a terminal symbol that comes from the NL input directly.
The example rule above is simplified for the convenience of explanation.
In practice, a rule often matches against generic variables rather than a specific word.
$S$ is the start variable that expands to other variables to match the whole query.

The grammar of the \flowsense semantic parser attempts to derive an input query by recursively searching for all possible matches (up to a preset limit) of the grammar rules.
This procedure is called {\it derivation}~\cite{BerantJ13}.
\flowsense uses the semantic parsing implementation from \sempre.
It also uses the Stanford CoreNLP~\cite{ManningCD14} toolkit that is built into \sempre for special utterance tagging.
The variables and rules (\ie \sempre formulas) are defined in \sempre grammar files.

\subsubsection{Special Utterance Placeholders}
The \flowsense grammar consists of static grammar rules and the special utterance placeholders.
The special utterance placeholders are at runtime dynamically replaced by their corresponding dataflow elements.
Therefore, the \flowsense semantic parsing is independent of the dataset, the dataflow diagram, and the analytical tasks.
The rules are generalizable across domains:
No new rules need to be created when the system switches to new datasets or tasks.

For example, \flowsense uses the generic variable $\var{column}$ in its grammar as a special utterance placeholder.
At runtime, a real column name (\eg ``mpg'') is automatically extracted from the dataset.
\flowsense identifies column names on the fly as the user types the query.
``mpg'' would show up as a tagged column, and then matched with $\var{column}$ by the parser.
A reverse mapping is performed from the placeholder to the particular column after query parsing so that the system may operate on that column.

Using special utterances in the grammar has several benefits.
First, special utterances enable \visflow functions to operate on elements that are important for dataflow diagram editing and visual data exploration.
Second, it makes the grammar set small as rules may be written with generic variables rather than specific dataset or diagram content.
Last but not least, the real-time tagging of special utterances provides important feedback to the user about what operations are available in the system and how the NLI interprets the query.

\subsubsection{Derivation Ambiguity}
\label{sec:derivation-ambiguity}
It is possible to have ambiguity when multiple possible query derivations exist, which can be defined as syntactic ambiguity~\cite{TongG15}.
For example, \flowsense uses wildcard variables to match general {\it table row} references.
Over the Auto MPG dataset, the token ``cars'' from \query{Show a plot of cars} describes the user's understanding of data entities but should be only treated as table rows from the NLI perspective.
Meanwhile, the token ``horsepower'' from \query{Show a plot of horsepower} is a special utterance and should be treated as a column to visualize.
Therefore a wildcard rule that matches ``cars'' as table rows may also match ``horsepower'', resulting in the second query getting improperly executed.
We could handle this case by creating a wildcard variable that rejects a special utterance token.
Nevertheless, such a design would lead to a larger number of variables and rules in the grammar, which are harder to maintain and develop.
Therefore we choose to resolve certain syntactic ambiguity in the parsing phase with supervised learning on a weight vector $\mathbf{w}\in \mathbb{R}^d$ that gives the probability of derivations based on input utterances.
Stochastic gradient descent (SGD) is employed to optimize the multiclass hinge loss objective~\cite{TaskarB03}, as introduced by Liang et al.~\cite{LiangP14} in the \sempre framework.
The objective is given by:

\begin{align*}
\min_{\mathbf{w}} \sum_{(x,y)}{\max_{y'}{\{\mathbf{w}\cdot{\textit feature}(x, y') + {\textit penalty}(y, y')\}} - \mathbf{w}\cdot{\textit feature}(x, y)}
\end{align*}


In the above, $x$ is the input query, $y$ is the preferred derivation, and $y'$ is a derivation choice.
The pair $(x, y)$ is iterated over all training data.
The feature of a derivation, {\it feature}$(x, y)$, maps the pair $(x, y)$ to a $d$-dimensional space and is determined by the applied rules in the derivation.
{\it penalty}$(y, y')$ is $0$ if $y = y'$ and $1$ otherwise.
The objective function has a penalty for possible choices of incorrect predictions that are within a margin of one from the correct predictions.
The parser fits the training examples by giving intended derivations higher probability so that they are preferred in case of ambiguity.
In particular, the rule that expands to a column special utterance will be preferred over a rule that expands to a wildcard.
Note that we only apply this training to facilitate the simplicity of the \flowsense grammar and reduce the number of required rules.
The training cannot address the ambiguity in natural language itself at large.
We were able to use a small training set of fewer than twenty examples to guide the preferred derivation in case of syntactic ambiguity for a rule set of around 500 rules.
This is feasible because the \flowsense rules are independent of data and dataflow diagrams.
The training set only needs to guide the semantic parser to focus on certain important grammatical features, such as special utterances or word proximity.


\subsection{Query Pattern}
The main goal of \flowsense is to support progressive construction of dataflow diagrams.
We studied the creation process of the \visflow diagrams in our sample diagram set and empirically identified a common pattern with five key {\itshape query components} that all \visflow functions may contain:
{\itshape function type}, {\itshape function options}, {\itshape source node(s)}, {\itshape target node(s)}, and {\itshape port specification}.
This pattern is illustrated in \autoref{fig:pattern} with a sample query \query{Visualize mpg, horsepower, and origin of the selected cars from MyChart in a parallel coordinates plot}.
In this query, the verb ``visualize'' implies applying a visualization function.
The three columns ``mpg, horsepower, and origin'' indicate the options (\ie what to visualize) for the visualization function.
The phrase ``from MyChart'' tells the system the location of the data to be plotted and provides source node information.
The phrase ``in a parallel coordinates plot'' indicates a new visualization node of the given visualization type is to be created as the target node.
As \visflow explicitly exports interactive data selection from visualization nodes, the phrase ``selected cars'' is a port specification that further describes that the user wants to visualize the selection from MyChart and the new visualization node should be connected to the selection output port of MyChart.

The grammar of \flowsense includes a variable hierarchy that matches the five key components of an NL query.
\autoref{fig:pattern} illustrates the parse tree that derives the sample query.
The variables involved in the derivation are shown in the parse tree, in which rule expansions are bottom-up.
A variable may carry information for multiple query components.
We design a broad set of variables and rules that are able to not only accept queries with a particular component order, but also their different arrangements.
For instance, ``Show mpg and horsepower in a scatterplot'' is equivalent to ``Show a scatterplot of mpg and horsepower''.
They both can be accepted by \flowsense.
\flowsense is also able to derive multiple functions from one single query and execute their combination, \eg \query{Show the cars with mpg greater than 15 in a scatterplot} infers both visualization and filtering functions.

A query may not necessarily contain all the five components explicitly.
For example, the user may simply say \query{Show mpg and horsepower} without mentioning any source node or target visualization type.
\flowsense may automatically locate source and target nodes in its query pattern completion phase (\autoref{sec:pattern-completion}).
An NL query may also contain implicit information, \eg \query{Find cars with maximum mpg} intends to perform data filtering to search for cars with the largest mpg value.
The use of a filter is identified by function classification in the query execution phase (\autoref{sec:function-classification}).


\begin{figure}[t]
\begin{center}
  \includegraphics[width=.9\linewidth]{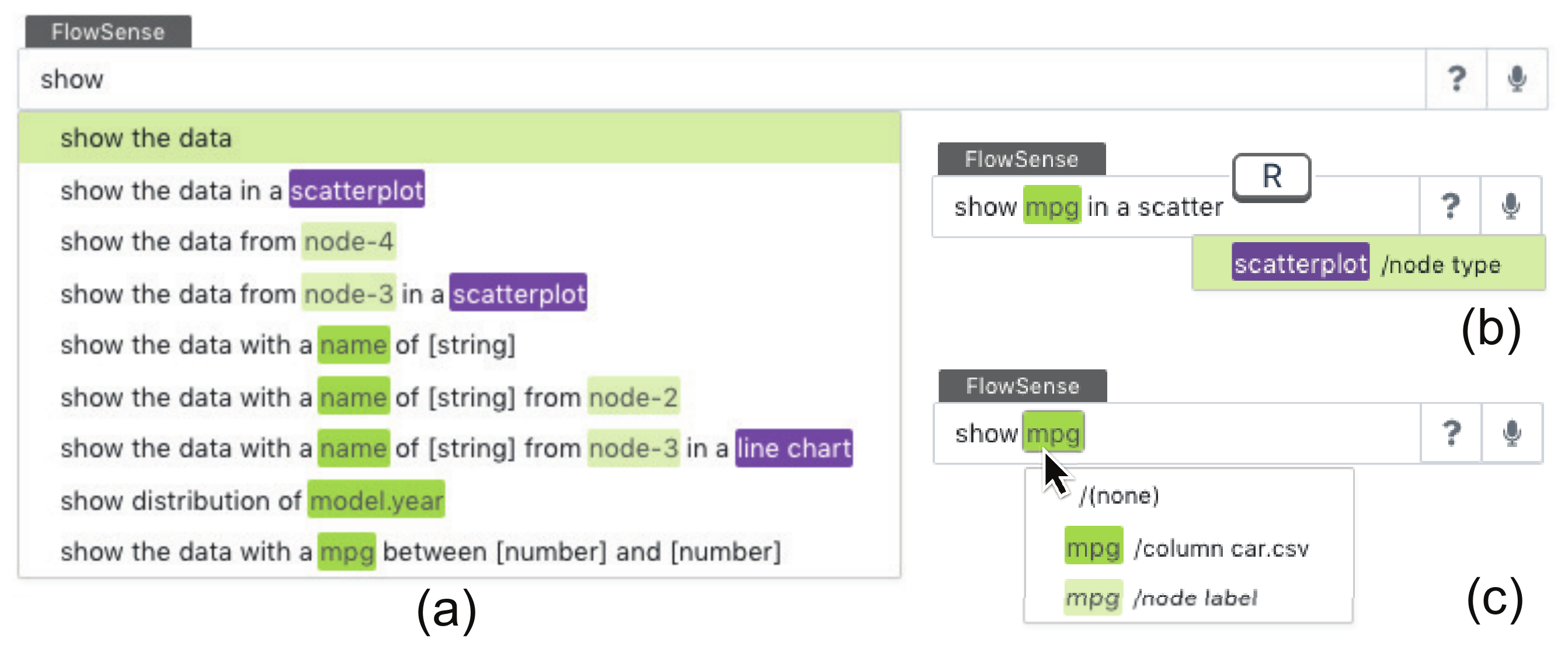} 
\end{center}
\vspace{-0.3in}
\caption{The \flowsense input box and its query and token auto-completion.
Special utterances are identified by unique colors. 
(a) Query auto-completion suggestions;
(b) Special utterance token completion: ``scatterplot'' is presented after the letter ``R'' is entered;
(c) Dropdown for handling tagging ambiguity: ``mpg'' are both column name and node label.
}
 \vspace{-5mm}
\label{fig:flowsense-ui}
\end{figure}

\subsection{Auto-Completion}
\label{sec:auto-completion}

The usability of an NLI is closely related to its discoverability.
It is desirable that when the query is partially completed, 
the system is able to provide hints or suggestions to the user about valid queries that include the partial input.
This has been a requested feature in prior NLI user studies~\cite{TongG15}.
We therefore develop an auto-completion algorithm in \flowsense to enhance its usability and discoverability.
When the user types a partial query and pauses, the system triggers query auto-completion automatically.
The auto-completion may also be invoked manually with a button press.
\autoref{fig:flowsense-ui}(a) shows the auto-completion suggestions in the \flowsense input box.

Auto-completion has been implemented in other visualization NLI, such as Eviza~\cite{SetlurV16}.
Eviza applies a template-based auto-completion, in which the system attempts to align user input to available templates.
Here we take a similar approach by creating a set of query templates with around one hundred queries.
Upon an auto-completion request, the algorithm searches through all possible textual matches between the user's partial query and a prefix of some template.
All matched queries are then sent to the \flowsense parser for evaluation.
If a query is accepted, it becomes an auto-completion candidate.
Some of the queries contain value placeholders and the user is expected to fill in those values ([string], [number] in \autoref{fig:flowsense-ui}(a)).

We also design a token completion algorithm that matches the partially typed word against available special utterances.
This helps speed up query typing with respect to the dataflow context.
The user may use the tab and arrow keys to select token completion candidates as in a programming IDE.
For example, when ``scatter'' is typed it can be completed to the available visualization type ``scatterplot'' (\autoref{fig:flowsense-ui}(b)).
Token auto-completion reduces typing workload and helps remind the user of the DFVS capability and the current dataflow diagram elements.


\section{Query Execution}
\begin{figure}[ht]
\begin{center}
  \includegraphics[width=\linewidth]{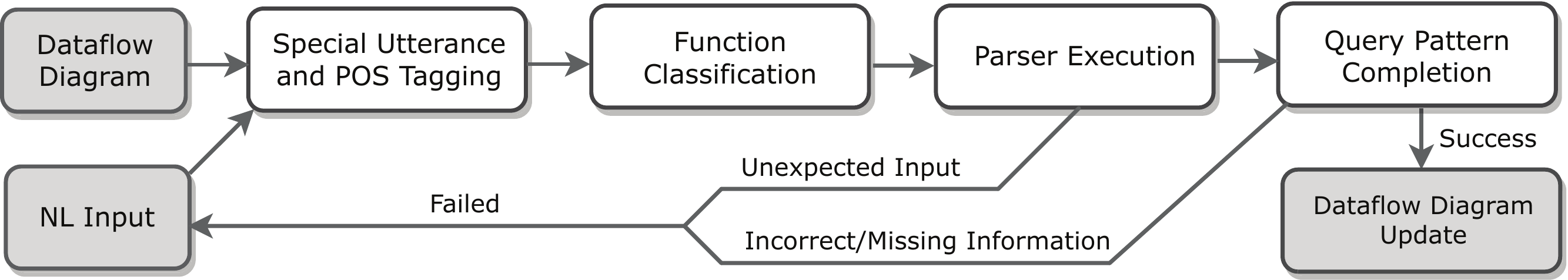} 
\end{center}
\vspace{-3mm}
 \caption{\flowsense query execution workflow.
 In case the grammar rejects the input, or there is no valid way to complete query components, a failure is returned to the user.
 }
 \vspace{-5mm}
\label{fig:exec}
\end{figure}

\flowsense is built as an extension to \visflow.
The user may activate the NLI at any time while working with the DFVS.
The user may either type the query in the input box or use the speech mode that is implemented with HTML5 web speech API.
In this section we introduce the query execution workflow as depicted by \autoref{fig:exec}.


\subsection{Special Utterance and POS Tagging}
\label{sec:tagging}
Special utterances have remarkable roles in executing a \visflow function.
Their tagging is performed on the fly when the user types the query.
For typo tolerance,
\flowsense employs approximate matching and checks each $k$-gram in the query (where $k$ may range from $1$ to the maximum special utterance word length) against all special utterances using case-insensitive Levenshtein distance~\cite{LevenshteinVI66,NavarroG01}.
We divide the distance over the string length and use the ratio to mitigate the fact that longer strings are more prone to typos.
We find a $k$ value of $2$ or $3$ and a ratio threshold of $0.2$ work well in practice.


In addition to recognizing special utterances, \flowsense also performs POS tagging on the query with CoreNLP.
Each token receives a POS tag as shown in \autoref{fig:pattern}.
POS tags are used to generalize the \flowsense grammar.
For example, many prepositions can be used interchangeably, \eg ``selection {\it of} the plot'' is equivalent to ``selection {\it from} the plot''.
Instead of having one rule for every preposition, the grammar uses a generic variable that matches any preposition.
POS tagging helps analyze the basic semantic structure of a query.



\subsection{Function Classification}
\label{sec:function-classification}
\flowsense uses keyword classification to identify the semantic meaning of words in the NL query and uses this information to decide a proper \visflow function to execute.
For instance, the verb ``show'' is a synonym of ``visualize'', ``draw'', etc.
These words indicate the intention to create a visualization.
Meanwhile, ``find'' may implicitly specify a data filtering requirement and is similar to ``filter''.
We compute the Wu-Palmer similarity scores~\cite{WuZ94} between words and use the measured scores to classify words in the NL query that have close meaning to a set of pre-determined \visflow function indicators.
The implementation of the similarity scores is based on WordNet~\cite{FellbaumC98} and NLTK~\cite{NLTK}.

\begin{figure*}[ht]
\begin{center}
  \includegraphics[width=0.9\textwidth]{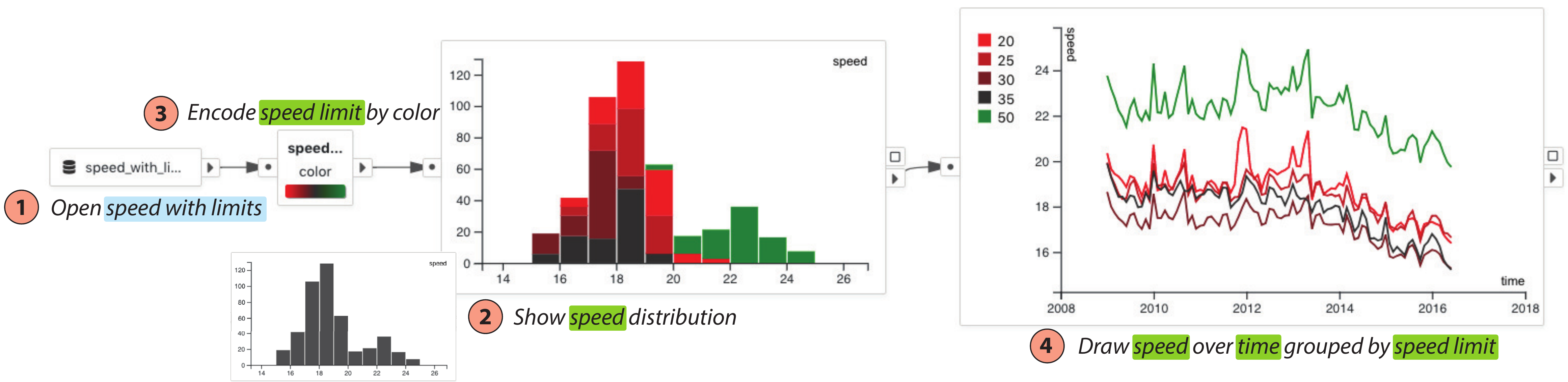} 
\end{center}
\vspace{-5mm}
\caption{
Using \flowsense to study the aggregated monthly average vehicle speed on NYC streets with different speed limits.
The queries are applied in the numbered order.
The result shows a histogram for speed distribution and a line chart for speed changes over time.
Both charts use color encoding based on the speed limit of the roads.
The smaller histogram snapshot shows the speed histogram without color encoding before step 3.
}
\vspace{-5mm}
\label{fig:visionzero-a}
\end{figure*}

\subsection{Query Pattern Completion}
\label{sec:pattern-completion}
After the parser identifies the existing key components of a query,
\flowsense attempts to fill in the blanks where information is missing using default values or the diagram editing focus.

\subsubsection{Finding Default Values}
Query components may be completed using default values.
Function options may have defaults.
For instance, \flowsense automatically chooses two numerical columns to visualize in a scatterplot triggered by a simple query \query{Show a scatterplot}.
Note that within a DFVS decisions like this can easily be changed by the user.
So \flowsense does not necessarily need to make a best guess.
Similar decisions include completing port specification.
By default \flowsense applies the newly created filter to all the data a visualization node receives, rather than the data subset interactively selected in the visualization.
Sometimes the default values may even be empty.
A query like \query{Filter by mpg} results in \flowsense creating a range filter on the mpg column with no filtering range given (\ie a no-op filter placeholder).
The user can then follow up and fill in the filtering range via the DFVS interface.

\subsubsection{Finding Diagram Editing Focus}
Whenever the user expands the dataflow diagram there always exists an editing focus, though often the focus is implicit.
For example, when the query contains a phrase like ``from MyChart'', the focus (\ie the source node of the query) is explicitly given.
However, users tend to neglect the source or target nodes in their queries, especially when there is a sequence of commands that together complete a task.
When a query does not have explicit focus,
\flowsense derives the user's implicit focus based on user interaction heuristics.
We compute a focus score for every node $X$ by:
\begin{equation}\nonumber
{\textit score}(X) = {\textit activeness}(X,t) + \alpha(1 - 
{
\frac{1}{1 + e^{-({\textit distanceToMouse}(X)/\gamma - \beta)}}
}).
\end{equation}

The activeness of $X$ is re-iterated upon every user click in the system:
\begin{align}\nonumber
{\textit activeness}(X,t) = {\textit activeness}(X,t-1) / 2 + {\textit click}(X, t),
\end{align}
where $\text{\it click}(X,t) = 1$ if the $t$-th click is on $X$ and $0$ otherwise.
This definition measures how actively a user focuses on a node by how many times she recently clicks on it, as well as how close it is to the mouse cursor.
The activeness derived from user clicks decreases exponentially over time, while the closeness to mouse dominates under a small distance with a shifted sigmoid function\footnote{See the appendix for more explanation on the characteristics of the diagram editing focus heuristics.}.
We find the parameters $\alpha = 2, \beta = 5, \gamma = 500$ achieve good result.
\flowsense chooses the node with the highest focus score to be the diagram editing focus. 
If multiple source nodes are required (\eg in a merge query), \flowsense selects the nodes in the order of their decreasing focus scores.

The focus may also be required by node type references. 
For instance, the user may input \query{show the data from the scatterplot}, in which ``scatterplot'' is a reference by node type that describes a scatterplot node existing in the dataflow diagram.
In case of a tie during the node type search, \eg there are multiple scatterplots in the diagram, the nodes with higher focus scores are chosen.

\subsubsection{Query Completion Ambiguity}
There may be multiple syntactically correct ways to execute a same query.
Consider the query \query{Show the cars with mpg greater than 15} applied on a visualization node.
From the grammar perspective the parsed outcome has no ambiguity: Apply an attribute filter and visualize the result.
However, there are two ways of execution:
One is to create a filter and then visualize the filtered cars in a new visualization;
Alternatively we may apply the filter on the input of the current visualization so that the existing visualization shows only the filtered cars.
Both can be desired under some circumstances.
\flowsense has the default behavior that prefers filtering the input when the source node is a visualization,
which we find empirically more intuitive.
Such ambiguity can often be resolved with a slightly refined query, \eg \query{Show the cars with mpg greater than 15 from the plot},
which would explicitly indicate that the filter should be applied to the output of the existing visualization.


\subsection{Diagram Update}
Once a query is successfully completed, \flowsense performs the \visflow function(s) with the given function options.
This typically results in the creation of one or more nodes, \eg the visualization function creates one plot while the highlighting function creates three nodes (Table~\ref{tab:functions}).
\flowsense may also update existing nodes without creating any new nodes, \eg when the user only changes rendering colors.
Additionally, a query may operate on multiple existing nodes at once, \eg linking and merging two tables create edges between two nodes.
Operating on multiple nodes together helps simplify dataflow interaction, as these operations previously require multiple drag-and-drop interactions.

After new nodes and edges are created, the diagram may become more cluttered.
\flowsense locally adjusts the diagram layout after each diagram update.
We use a force-directed layout modified from the D3 library~\cite{D3} that manipulates the vicinity of the current diagram editing focus. 
We extend the force to take rectangular node sizes into account so that larger nodes such as embedded visualizations have stronger repulsive force for avoiding node overlap.
User-adjusted node positions are remembered by the system,
and the layout algorithm avoids moving nodes that have been positioned by the user. 
Currently \flowsense does not look for an optimal dataflow layout.
We leave layout improvement~\cite{BatiniC86} for future work.


\subsection{Error Recovery}
\label{sec:error-recovery}
There are several types of potential errors in executing a query:

\noindent (1) The query cannot be accepted by the grammar.
For example, out-of-context input (\query{What time is it now}) and unsupported functionality (\query{Split the data into two halves}) would receive grammar rejection;

\noindent (2) The query is grammatically correct but invalid based on the dataflow context, possibly due to incorrect references of dataset and diagram elements.
For example, the user may attempt to show data from a non-existing node, \eg asking to \query{Highlight the selected cars from the scatterplot} when there is no scatterplot in the dataflow.
Such errors are captured at the query pattern completion step.

\noindent (3) The query is executed fully but does not meet the user's expectation.
For example, \query{Show the data} by default creates a scatterplot but the user instead wants a heatmap, or \query{Merge these two nodes} merges an unexpected pair of nodes when ``these'' appears to be a vague reference (the system chooses two nodes with the highest focus scores).

Upon the first two types of errors the system displays a message and asks for a query correction.
For the last type of error it is up to the user to adjust the dataflow diagram.
Since the user is simultaneously using the underlying \visflow DFVS while using \flowsense,
she always has the flexibility to undo the \flowsense action or to make partial adjustments when the NLI does not yield exactly the desired outcome.


\section{Evaluation}
\label{sec:evaluation}

To evaluate the effectiveness of \flowsense, we describe the results of one case study and one formal user study.

\subsection{Speed Reduction Study}
We invite several users to try out the \flowsense prototype in different data analysis domains and analyze their usage of our NLI.
In this paper we introduce one case study in which we work with two domain experts in person to address a practical research task using a comprehensive set of NL queries.
The analysts are researching the city regulation issued on November 7, 2014
that reduces the default speed limit on all New York City streets from 30 MPH to 25 MPH.
The data contain the estimated average hourly speed~\cite{PocoJ15} for each road segment in Manhattan from January 2009 to June 2016.
The speed estimation was performed based on the TLC yellow taxi records~\cite{TLC} that only have pickup and dropoff information.
The analysts are familiar with the data, and the visualizations to be created are similar to the visualizations they previously generated for the project using Tableau~\cite{tableau}.
However they have no prior experience with either \visflow or \flowsense.
We met the analysts in person and first introduced \visflow and \flowsense in a 30-minute session.
Then we guided the analysts through how \flowsense can be used to create visualizations to study the speed reduction.
We observed in this study that almost all the analysts' visualization requests (excluding those that exceed the scope of the \visflow subset flow) can be effectively supported by \flowsense.
Here we summarize the NL queries applied in the speed reduction study.

Initially, the analysts would like to look at the speed reduction impact at a larger scale.
They first load a pre-computed speed table (\autoref{fig:visionzero-a}(1)) with the \flowsense data loading utility function (the analysts know the dataset name).
The table contains the monthly average speed aggregated by the speed limits of the streets.
The analysts ask the system to present a histogram of speed by \query{Show speed distribution} (\autoref{fig:visionzero-a}(2)).
The first histogram has no color encoding but the analysts are able to immediately add a color scale by \query{Encode speed limit by color}.
\flowsense inserts a color mapping node with a red-green color scale at the input of the histogram (\autoref{fig:visionzero-a}(3)).
The histogram shows the street groups with higher speed limit in green, and lower speed limit in red.
To view the speed changes over time, the analysts use the query \query{Draw speed over time grouped by speed limit} (\autoref{fig:visionzero-a}(4)).
The query result is a line chart showing average speed changes for different speed limit groups.
The analysts observe that overall there is a speed reduction in all speed limit groups that started around middle 2013.

Seeing the overall trend, the analysts move on to a comparative analysis between individual streets from two slow zones.
They load and visualize a table about speed limit sign installation in a map (\autoref{fig:visionzero-b}(1)) by \query{Show the data in a map}.
This dataset has for each road segment in Manhattan its speed limit, geographical location, and whether the street has speed limit signs installed (signs are shown as dots in the map).
As the slow zones mostly have speed limit signs installed, the analysts narrow down the data in the map by placing a filter on the ``sign'' column (\autoref{fig:visionzero-b}(2)).
The filtered map reveals two slow zone neighborhoods with densely located signs: Alphabet City and West Village.
The analysts apply one map visualization for each zone for a comparison between the two zones.
They label the two maps by the slow zone names and select a few streets from each zone (marked in the maps of \autoref{fig:visionzero-b}).
To study the speed changes of these selected streets, another table (named ``segment monthly speed'', also known to the analysts) that includes monthly average speed for each road segment is added to the diagram (\autoref{fig:visionzero-b}(3)).
The analysts then use the link queries to create a sequence of nodes that extract segment IDs from the selected streets and find their monthly average speed from the segment monthly speed table (\autoref{fig:visionzero-b}(4)).
Blue and red colors are assigned to the streets in West Village and Alphabet City respectively to visually differentiate them (\autoref{fig:visionzero-b}(5)).
The two groups of streets are then merged by a subset manipulation function (\autoref{fig:visionzero-b}(6)).
Note that the query \query{Merge} only has a single word.
It works because the query completion of \flowsense automatically locates the recently focused color editors as the source nodes for this query.
Finally, the two groups are rendered together in a speed series visualization (\autoref{fig:visionzero-b}(7)), which compares the speed changes between the two groups of streets.
As the visualizations produced by \flowsense are linked,
the analysts can easily change the street selection in the maps to compare different groups of streets.

This case study demonstrates that \flowsense can be applied to a practical, comprehensive analytical task.
The generated visualizations may guide the analysts towards further data analysis.
The analysts participating in this study think \flowsense is helpful,
especially since it exemplifies how to build \visflow diagrams and facilitates their learning of the DFVS.

\subsection{User Study}
\label{sec:user-study}

We conduct a formal user study to evaluate the effective of \flowsense together with the \visflow framework.
Through the user study we validate whether a user is able to smoothly apply \flowsense for dataflow diagram construction, and how well \flowsense responses meet the user's expectation.
We design an experiment that introduces \flowsense and \visflow to the participant and assigns analytical tasks to be solved within the system.

\subsubsection{Experiment Overview}

The user study is carried out in a fully automated manner using an online system with step-by-step instructions.
The participants join the study using a web browser on their own machines.
Participants may ask the experiment assistant for help and clarification via web chat or phone call during the experiment session.

We recruited $17$ participants ($11$ male, $6$ female, all with an age between $20$ and $30$) who work or study in the field of computer science.
$12$ participants have a data visualization background.
$9$ are graduate students, and the other $8$ are professionals (software engineer, researcher, faculty).
$3$ participants have prior experience with \visflow.
No participants have prior knowledge about \flowsense.
The participants are chosen to have a variety of specialities so as to represent potential DFVS users.
The participant group includes visualization designers, data scientists, and software engineers who share data analysis interest but have different skill sets.
The study is structured into two phases:

\myparagraph{Tutorial Phase}.
The participant completes a tutorial of the \visflow dataflow framework, and then a tutorial of the \flowsense NLI.
After each tutorial, the participant is asked to complete the tutorial diagram to demonstrate familiarity with the introduced tool.
Each tutorial is expected to take 10 to 20 minutes.
After the tutorials there is an on-demand practice session with a flexible duration.

\myparagraph{Task Phase}.
The participant explores an {\it SDE Test} dataset and constructs dataflow diagrams using \flowsense and \visflow to answer questions about the data.
The participant is encouraged to use \flowsense as much as possible.
The usage of the NLI is not enforced because the goal of the NLI design is to improve the user experience of the DFVS, rather than to completely replace the traditional DFVS interactions (which is likely infeasible).
The entire task phase is expected to take 30 to 60 minutes.

At the end of the study, the participant takes a survey to give comments and quantitative feedback about \flowsense and \visflow.

\subsubsection{Dataset and Tasks}

The {\it SDE Test} dataset includes the test results of software development engineer (SDE) candidates stored in two tables. 
The first table describes the test results for each candidate.
A test consists of answering several multi-choice questions selected by the system from a large question pool.
Each question has a unique ID, a pre-determined difficulty, its supported programming language(s), and possibly a time limit.
For each question, the candidate receives a result (correct, wrong, skipped, unanswered)\footlabel{remarks}{See the appendix for additional remarks and results of the user study.}.
The dataset also has a ``TimeTaken'' column that stores how much time a candidate took to answer a question.
The second table includes background information about each candidate, such as the candidate's highest degree level, field of study, and institution.
We give three analytical tasks about this dataset.
The tasks are designed to reflect common tasks performed in visual data exploration:

\myparagraph{(T1) Overview Task}.
The participant is asked to visualize the overview distribution of the question answering results, and figure out the total number of questions that were skipped, and the percentage of a question being answered correctly.

\myparagraph{(T2) Outlier Task}.
The participant is first asked to find a candidate with an outlier background information value (who incorrectly entered the current year ``2018'' in place of his own information).
Then the participant is asked to investigate a data recording discrepancy regarding the ``TimeTaken'' column:
Some of the ``TimeTaken'' values are erroneously large numbers when a question is unanswered.

\myparagraph{(T3) Comprehensive Task}.
The participant is asked to identify one question that Masters candidates answer significantly better than Bachelors candidates.
This task requires comprehensive usage of the dataflow features: attribute filtering, brushing, and heterogeneous table linking.

All the three tasks have definitive correct answers to ensure that participants explore the data and draw conclusions reasonably.
Each user study session is logged with anonymous full diagram editing history.
We analyze the study results based on task answers and completion time, comments and quantitative feedback, and NL query logs.

\subsubsection{Task Completion Quality}

\begin{figure}
\centering
\subfigure[]{\label{fig:a}\includegraphics[width=.9\linewidth]{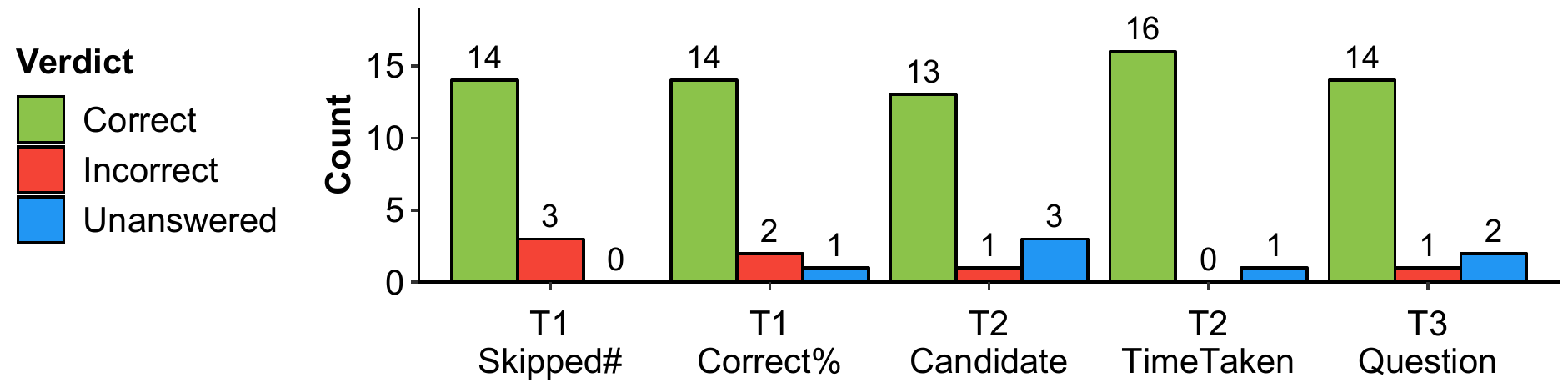}}
\addtolength{\subfigtopskip}{-10pt}
\subfigure[]{\label{fig:b}\includegraphics[width=.9\linewidth]{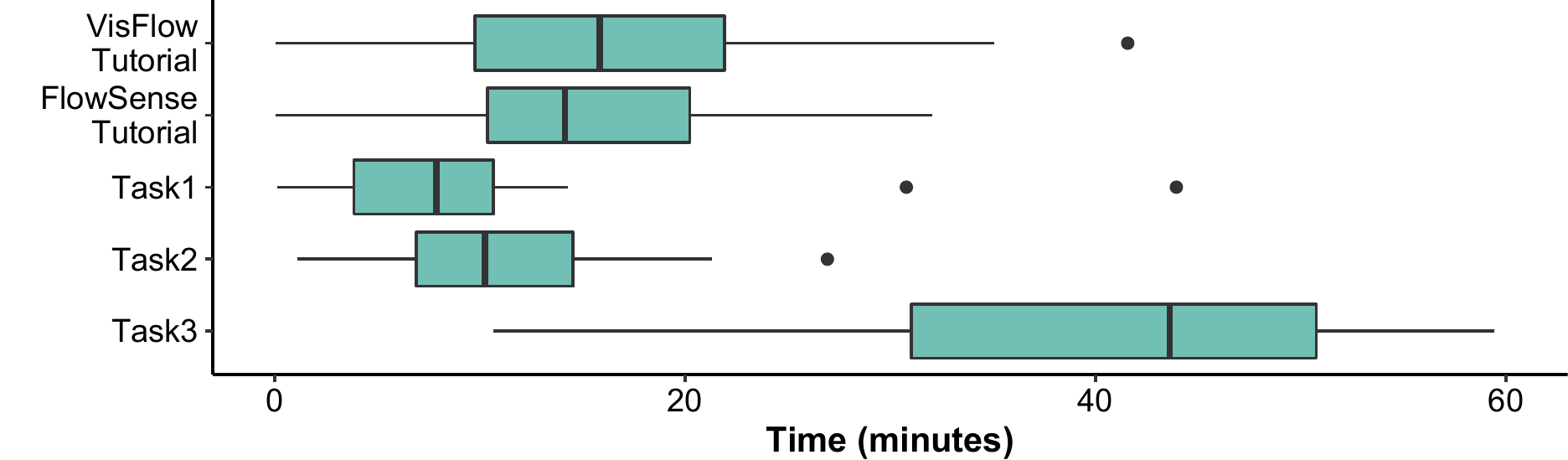}}
\vspace{-3mm}
\caption{
(a) Verdict distribution of participant answers to each of the user study tasks.
(b) Box plot of completion time for each user study step\protect\footref{remarks}. 
}
\vspace{-5mm}
\label{fig:answers}
\end{figure}

\autoref{fig:answers}(a) shows the verdict distribution of the participants' answers.
It can be seen that the majority of the participants were able to come up with the correct answers to the tasks.
\autoref{fig:answers}(b) shows the completion time distribution for each step of the user study.

It can be observed that the time taken for the tutorials and tasks are mostly as expected.
Yet the time required for a task increases when the task involves heterogeneous tables and interactive data filtering to find solutions (T3).
After reading the user comments in the feedback, we believe this may be due to the fact that many participants are first-time \visflow users and need to digest the concept of the \visflow subset flow model.
In particular, linking heterogeneous tables can be challenging to understand at first.
However, most users were able to get the idea and formulate a solution.
This is reflected by one of the feedback comments:
\query{The linker functions are confusing at first. But after experimenting with the tool for a while and getting to know how they work, things become easier.}
We believe such a learning curve is natural for DFVS.

\subsubsection{Quantitative Feedback}

We ask for feedback on six aspects regarding \flowsense (and also VisFlow\footref{remarks}) in our survey.
Each aspect is presented with a statement and a $1$--$5$ Likert scale for the participant to express agreement ($5$) or disagreement ($1$).
Table~\ref{table:flowsense-survey} lists the feedback for the \flowsense NLI.
The quantitative feedback shows that most users were able to understand the scope of \flowsense,
and apply it for dataflow diagram construction.
The users were also asked to compare the NLI-assisted dataflow usage against their earlier experience in the tutorial phase with the standalone \visflow framework.
Twelve users agree (with a feedback score of at least 4) that \flowsense simplifies the diagram construction, and ten users agree that \flowsense speeds up the data exploration.

The feedback also reveals space for improving the NLI.
In particular, it is unclear to most users how to update a rejected query to make it accepted.
It may be helpful to design an algorithm that provides suggested corrections or changes to a failed query.
However, this is technically challenging as changing minimally a query to fit it into the parse tree is algorithmically non-trivial.
We would like to leave query correction suggestions for future work.

\subsubsection{Query Log Analysis}

To closely study where \flowsense does not accept a query, we manually went over the rejected queries and categorized each rejected query by its reason of rejection.
Overall, we analyzed $649$ queries, out of which $421$ were accepted by \flowsense. 
Excluding the $34$ invalid and mistyped queries, the raw acceptance rate was $68.455\%$. 
We found some of the rejection issues straightforward to resolve: the requested functionality was not implemented, bugs in the query execution code, etc.
We were able to fix those issues in a short iteration of the NLI implementation, resolving $34$ ``not implemented'' queries and $18$ software bugs.
The improved acceptance rate would be 76.911\%.
In general, it requires systematic engineering efforts to thoroughly increase query coverage for the ``not implemented'' category, which is beyond the scope of this paper.
The remaining unresolved failures are summarized in \autoref{fig:reasons} with their counts\footnote{See the appendix for the detailed definition and examples for each category.}.


\begin{table}[ht]
\def\arraystretch{0.6}
\setlength\tabcolsep{2pt}
\begin{tabular}{ p{.6\linewidth}  p{.27\linewidth} }
\toprule
{\bfseries \makecell{\small Aspect}} & {\bfseries \makecell{\small Feedback} }\\
\midrule
\small I understand what queries FlowSense may accept and execute. & \raisebox{-.6\height}{\includegraphics[width=\linewidth]{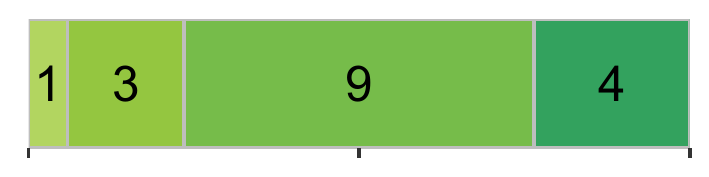}} \\
\small The responses of FlowSense meet my expectations. & \raisebox{-.6\height}{\includegraphics[width=\linewidth]{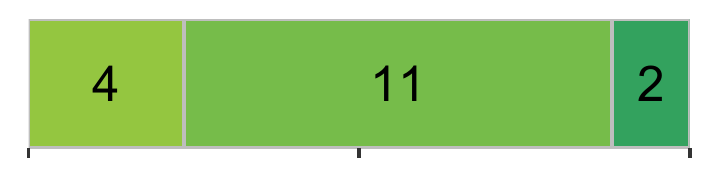}} \\
\small FlowSense simplifies dataflow diagram construction. &  \raisebox{-.6\height}{\includegraphics[width=\linewidth]{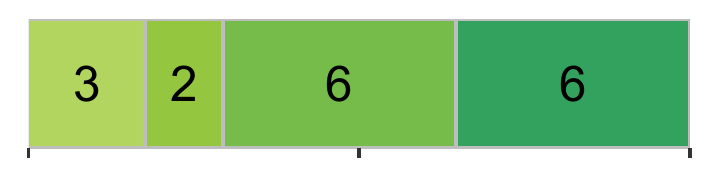}} \\
\small FlowSense speeds up my data exploration. &  \raisebox{-.6\height}{\includegraphics[width=\linewidth]{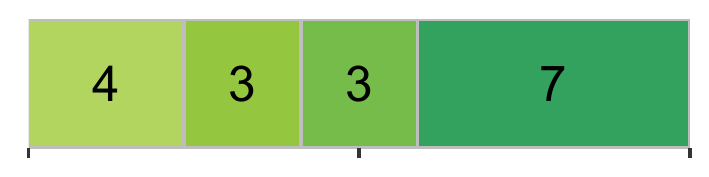}} \\
\small FlowSense helps me learn VisFlow features that I was not aware of. & \raisebox{-.6\height}{\includegraphics[width=\linewidth]{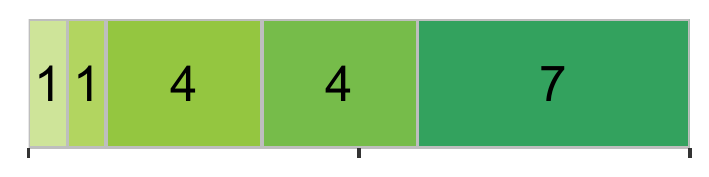}}\\
\small When my query got rejected, I can figure out how to update it to let it be accepted. & \raisebox{-.6\height}{\includegraphics[width=\linewidth]{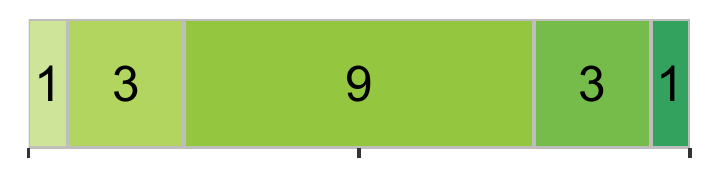}} \\
\bottomrule
\end{tabular}
\includegraphics[width=0.09\linewidth]{study/scores}

\vspace{2mm}
\caption{FlowSense Survey Result.
The feedback column shows the score distribution for each assessed aspect of the NLI.
The numbers on the colored bars show the counts of the scores received.
Darker green represents higher score.
}
\vspace{-6mm}
\label{table:flowsense-survey}
\end{table}

Some of those failures are more challenging to resolve.
Specifically, \flowsense does not make logical inferences and deals only with the raw values in the data.
If the user rephrases the query by natural language variation or implication (26 occurrences in~\autoref{fig:reasons}), the query would be difficult to parse.
The query \query{Show only segments with signs} is more natural than that in \autoref{fig:visionzero-b}(2).
Yet \flowsense does not infer that a segment with a ``sign'' value of ``yes'' implies that it is a segment ``with sign''.
In T3 the dataset has ``HighestLevelOfEducation'' as a column name, but if the user mentions ``degree'', \flowsense does not know that it is equivalent.
There needs to be additional knowledge base added to the system so that the NLI can detect concept equivalence, which is generally difficult to achieve.
In a ``composite'' query, the user intends to perform several \visflow functions in one query (\eg creating nodes, applying filter, and assigning color together).
It is difficult to write concise grammar rules to accept composite queries.
In practice, by informing the users of these limitations, in most cases the issues can be circumvented via rephrasing the queries,
\eg composite queries can be split into smaller steps that are easier to parse and execute.

When an operation requested is not supported by the DFVS, a ``not supported'' failure arises,
\eg \visflow without its data mutation extension\footnote{https://visflow.org/extension. The extension is not yet supported} cannot aggregate and mutate data.
When the special utterance tagging over-aggressively tags a non-special word, its placeholder fails to resolve, leading to a ``tagging error''.
The user may use the token dropdown in the \flowsense input box to correct tagging mistakes,
or disambiguate tokens with multiple meanings (\autoref{fig:flowsense-ui}(c)).

\begin{figure}[htbp]
\begin{center}
  \includegraphics[width=.95\linewidth]{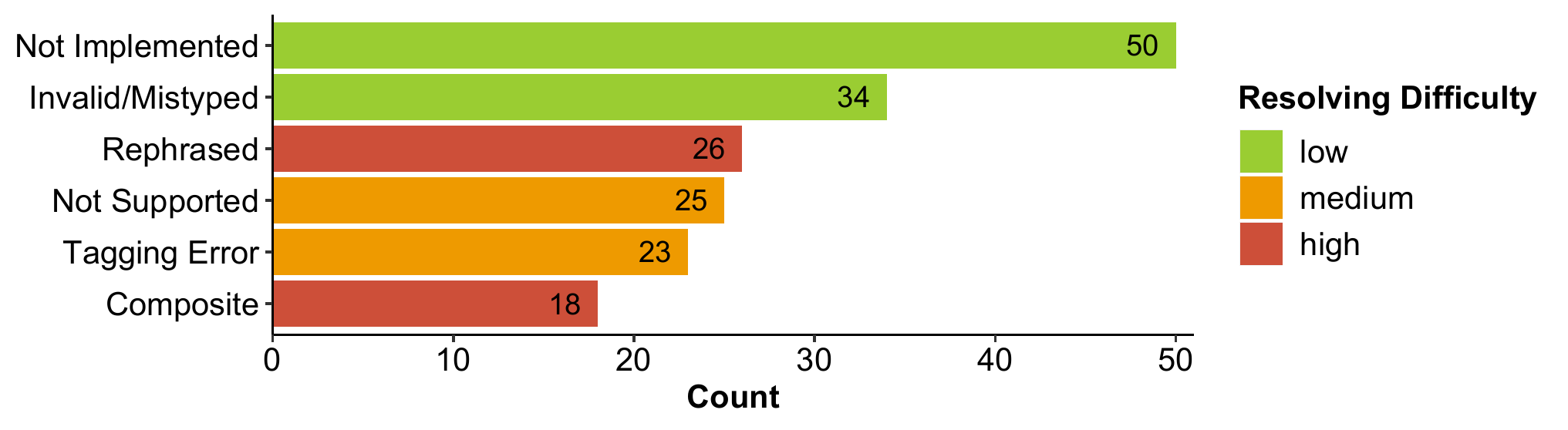}
\end{center}
\vspace{-7mm}
\caption{Number of failed queries grouped by the reasons of their failures.
The colors of the bars indicate the relative difficulty of resolving a failure.}
\vspace{-3mm}
\label{fig:reasons}
\end{figure}

\section{Discussion and Limitations}
\label{sec:discussion}


\subsection{Scalability and Generalizability}
Technically there are many ways to create a set of rules that implement the same dataflow function.
Following the active development and enhancement of the system functionality, from time to time grammar rules can be combined and rewritten to make the grammar more concise.
We keep iterating and refining the \flowsense grammar to expand its functionality.
\flowsense currently includes about 200 variables and a rule set of around 500 rules in its grammar.
Our grammar development practice employs continuous integration and maintains a test set (currently of 131 test queries) to ensure that all categories of \visflow functions may execute properly during iterations and extensions of the grammar.
Approximately 10 to 20 rules need to be added to support a new dataflow function category.

Though the grammar rules of \flowsense are coupled with the underlying \visflow functionality,
its approach of utilizing special utterance placeholders is generalizable to other dataflow systems that employ similar modular component design.
Once the data- and diagram-independent dataflow elements are identified,
these elements can be represented by special utterances in the grammar and dataflow implementation can subsequently be extended to process them.
For example, we may extend the grammar to support more data processing power obtainable from a computational dataflow system like KNIME~\cite{Knime}.

\subsection{User Behavior and Engagement}
The effectiveness of a grammar-based semantic parser couples with the grammar design.
One design flaw in the grammar may result in unexpected rejections of seemingly acceptable queries.
Despite careful grammar design, the user is likely to come up with questions that exceed the scope of the grammar.
However, we find that users are willing and able to refine rejected queries with a small number of trial-and-error attempts.
Besides users may become more proficient with the NLI after reading query examples so as to understand the NLI capability.
Yet showing too many examples may limit the user's thoughts and forfeit the benefit of using an NLI.
We would like to further study user behavior regarding NLI usage in DFVS in the future to better identify when and what query examples need to be provided.

We also observe that users tend to perform composite queries and ask for batch operations using the NLI.
With traditional mouse/keyboard interaction, the results of such queries have to be achieved by a sequence of interactions.
\flowsense increases the data exploration efficiency by naturally enabling batch operations.
In fact, we notice some users were able to repeat successful short queries that achieved the most batched result.
The convenience of using NL to carry out multiple operations may improve the user's engagement~\cite{Cohen92},
provide interaction ``shortcuts'',
and make dataflow features more accessible by simplifying the creation of rather complicated sub-diagrams, \eg ``highlighting''.

\subsection{Technique and Scope}
We prefer semantic parsing to deep learning mainly because the latter has a bottleneck of requiring a large volume of training examples.
Though there are benchmark datasets for general NLP, there has not yet been a training set catered for visualization-oriented NLI or DFVS.
In the future with more users working with the NLI, we may collect more user queries that constitute a rich training set in order to support methods like neural networks for text classification~\cite{YoungT18}.

Currently \flowsense only works with dataflow diagram editing.
But it may be desirable for the NLI to answer analytical questions such as \query{Does the vehicle speed decrease over years in NYC?} by creating a visualization like \autoref{fig:visionzero-a}(4).
To that end we need further research on the dataflow functions and their application to answering analytical questions.
One possible direction is to study how DFVS diagrams can be constructed for knowledge-based Q\&A~\cite{ClarkP99}.


\section{Conclusions}
\label{sec:conclu}

In this work we design \flowsense, a novel NLI for visual data exploration within a DFVS.
We build \flowsense for the \visflow framework and show that it improves the DFVS usability and simplifies diagram construction.
\flowsense applies semantic parsing to map NL input to \visflow functions.
Its emphasis on special utterances and usage of special utterance placeholders make the semantic parsing independent of datasets and diagrams, but at the same time aware of the dataflow context.
The real-time feedback of tagged special utterances, as well as query and token auto-completion features, largely helps the user understand the underlying parsing state.
Our case study and user study results demonstrate the effectiveness of the proposed NLI, and help identify future research directions for its improvement.

\acknowledgments{
%
We would like to thank BlindData.com for providing the user study dataset.
This work was supported in part by: the Moore-Sloan Data Science Environment at NYU;
NASA; NSF awards CNS-1229185, CCF-1533564, CNS-1544753, CNS-1730396, CNS-1828576. 
B. Yu and C. T. Silva are partially supported by
the DARPA MEMEX and D3M programs. Any opinions, findings, and conclusions or 
recommendations expressed in this material are those of the authors and do not necessarily 
reflect the views of DARPA.
}

\bibliographystyle{abbrv}

\bibliography{../main}
\end{document}



\appendix

\section{FlowSense Grammar Design}
We provide an open source repository that contains the details of the \flowsense implementation: \url{https://github.com/yubowenok/flowsense}.
This repository includes the grammar rules, backend API (implemented in TypeScript and Python), and integration tests.
The structure of this repository and its installation and setup guide can be found within its \href{https://github.com/yubowenok/flowsense/blob/master/README.md}{README} file.

In particular, the grammar rules are located in the \href{https://github.com/yubowenok/flowsense/tree/master/src/grammar}{{\tt *.grammar} files}.
The entry point is {\tt main.grammar}.
The grammar rules are written in the \sempre grammar format (\ie SEMPRE formulas).
More details on SEMPRE can be found at the \href{https://github.com/percyliang/sempre}{\sempre GitHub repository}.

\section{Characteristics of the Diagram Editing Focus Heuristics}
Intuitively, the focus score keeps track of the diagram element that is last interacted with.
It has two components: the activeness resulted from mouse clicks, and the distance-to-mouse bonus.
The activeness score exponentially decreases when there is no interaction on the node, while the distance-to-mouse bonus prioritizes the elements around the last interaction.

When the mouse hits a node $x$, node $x$ receives a high activeness score of one from the {\it Click(X, t)} part,
which almost certainly ensures that the focus score of $x$ is higher than any other node $y$ that is not interacted with.
Though $y$ (when it is in the proximity of $x$) may receive a distance-to-mouse bonus that remedies its exponential loss on the activeness score, 
note that $x$ receives a distance-to-mouse bonus too, and the bonus can only be higher than the bonus received by $y$ because $x$ is clicked on and thus closer to the mouse.
Therefore, the outcome is that $x$ becomes the first prioritized node, and $y$ becomes the second prioritized.
In other words, if a \visflow function requires two node operands, then $x$ is chosen first, and then $y$ is chosen.

If the user clicks on the background, all nodes have exponentially decreasing activeness score, 
and their distance-to-mouse bonus will likely dominate the focus score.
Consequently, the nodes that are closer to the last click become the chosen query targets.
As there can be multiple nodes around the background click,
occasionally a node not actually focused by the user may happen to be close to an unintentional background click (\eg accidentally performed during canvas panning).
The next NL query may then be incorrectly performed on this node.
This error can be fixed by clicking on a specific node to focus on it and redoing the NL query.

\section{Additional User Study Remarks}
\begin{enumerate}[*]
\item In the SDE test, answering a question wrong results in negative score penalty. 
Therefore skipping a question can be worthy.
Skipping requires an explicit button click.
The ``unanswered'' result is given when the user has no action within the allocated time limit of a question.
\item In Fig. 6(b), four outliers due to interruptions on the participant's end are not shown: 2550 minutes on Task1,  and 109, 119, 212 minutes on Task3 were measured as the task completion time that includes the interruptions.
\end{enumerate}


\begin{table}[h!]
\def\arraystretch{0.6}
\begin{tabular}{ p{.6\linewidth}  p{.25\linewidth} }
\toprule
{\bfseries \small \makecell{Aspect}} & {\bfseries \small \makecell{Feedback} }\\
\midrule
\small I understand the majority of VisFlow features. & \raisebox{-.6\height}{\includegraphics[width=\linewidth]{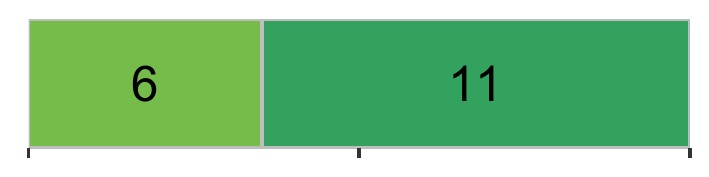}} \\
\small I understand the subset flow in VisFlow. & \raisebox{-.6\height}{\includegraphics[width=\linewidth]{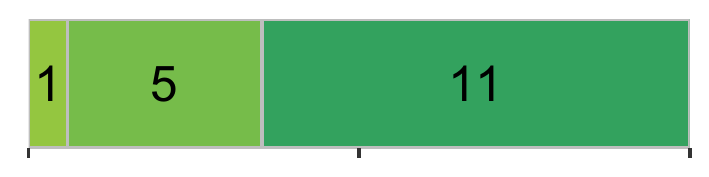}} \\
\small I can follow the VisFlow dataflow diagram and understand their functionality. & \raisebox{-.6\height}{\includegraphics[width=\linewidth]{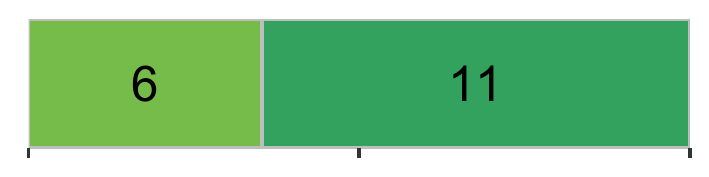}} \\
\small VisFlow is relatively simple to learn and use. & \raisebox{-.4\height}{\includegraphics[width=\linewidth]{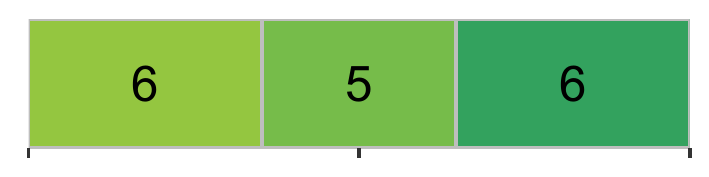}} \\
\small VisFlow is an effective system for visual data exploration. &  \raisebox{-.4\height}{\includegraphics[width=\linewidth]{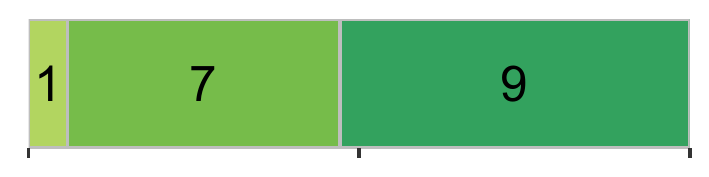}} \\
\small I would like to use VisFlow for my future data exploration tasks. & \raisebox{-.4\height}{\includegraphics[width=\linewidth]{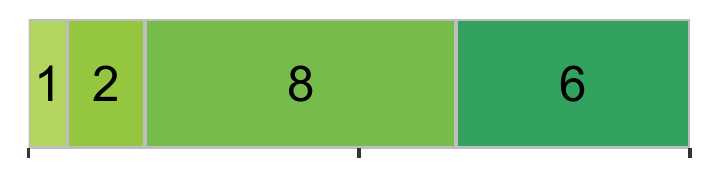}} \\
\\ \bottomrule
\end{tabular}
\includegraphics[width=0.07\linewidth]{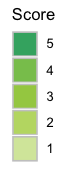}
\caption{VisFlow Survey Result.
The feedback column shows the score distribution for each assessed aspect of \visflow.
The numbers on the colored bars show the counts of the scores received.
Darker green represents higher score.
Overall the users were able to understand well the DFVS functionality and use it effectively for visual data exploration.
}
\label{table:visflow-survey}
\end{table}

\section{VisFlow Survey Results}

The \visflow user study is part of the \flowsense user study.
Before \flowsense was introduced to the users, a tutorial on the details of the \visflow dataflow framework was given.
The participant group of the \visflow user study is thus exactly the same as the group described in the paper:
In total, $17$ users ($11$ male, $6$ female, all with an age between $20$ and $30$) who work or study in the field of computer science participated in this study.
$12$ participants have a data visualization background.
$9$ are graduate students, and the other $8$ are professionals (software engineer, researcher, faculty).
$3$ participants have prior experience with \visflow, who may yet formally evaluate it through task completion.
The participants are chosen to have a variety of specialities so as to represent potential DFVS users.
The participant group includes visualization designers, data scientists, and software engineers who share data analysis interest but have different skill sets.

The users were given a form to assess the effectiveness of \visflow quantitatively using a Likert scale of 1 to 5 (5 is ``strongly agree'' and 1 is ``strongly disagree'').
\autoref{table:visflow-survey} shows the quantitative survey feedback for the \visflow DFVS.
It can be observed that the users were able to understand the subset flow model of \visflow.
The majority of the users agree (with a score of at least 4) that \visflow presents an effective approach to visual data exploration,
and can successfully utilize \visflow features for their data exploration.

\section{Query Analysis -- Failure Category Description}

This following list provides the description of each query failure category we identified in the user study results:

\begin{itemize}
\item \myparagraph{Not Implemented}.
\flowsense grammar may technically support parsing this query.
Yet we have not implemented the corresponding grammar and its web client handler (query execution code for diagram update).
Example queries include \query{change the x column to mpg}.
The current system implementation does not support node option changes triggered by the NLI (except for visual editors).
Queries of this category can be accepted by extending the grammar and adding more rules.

\item \myparagraph{Invalid/Mistyped}.
The query is an invalid sentence and cannot be understood by a human;
Or the query has mistyped words and fails to describe the intended data entity or dataflow element.

\item \myparagraph{Rephrased}.
The user rephrases the query using grammatical structures not expected by the grammar, 
or the user uses words that do not appear in the dataset table to describe a table column or cell value.
For example, in Task 3 if the user mentions ``degree'', \flowsense does not know that ``degree'' is equivalent to the ``HighestLevelOfEducation'' column in the data.
Though one can inform a system of such equivalence case-by-case or find synonyms from WordNet,
it is non-trivial to generally detect such equivalence.
Consider the equivalence between \query{Show only segments with signs} and \query{Show only segments with a sign of yes}: 
``yes'' is not an immediate synonym of ``with'' and their common implication of ``existence'' is subtle.
There needs to be additional knowledge base added to the system to support the detection of concept equivalence.

\item \myparagraph{Not Supported}.
The functionality indicated by the query is not supported by the \visflow dataflow framework.
This is not an issue of the NLI but a limitation of the underlying DFVS.
A query like \query{How many questions were skipped} asks directly an analytical question about the dataset and exceeds the scope of \visflow.
It cannot be accepted by simply extending the grammar because there needs to be a reasonable way to construct dataflow sub-diagrams to answer the analytical questions,
which can be complex and challenging to identify.

\item \myparagraph{Tagging Error}.
A special utterance should have (have not) been tagged, but it was not (was) tagged.
For example, the query \query{Select iris with id between 3 and 5} has the word ``iris'' that is both a word to describe the data entity and a dataset name.
When \flowsense automatically tags ``iris'' as a dataset name special utterance, the parser may fail to accept the query.
In this case the user may manually override the tagging to avoid the error resulted from parsing ambiguity.
In future work we may also explore techniques that can be integrated into the parser to structurally reduce such errors.

\item \myparagraph{Composite}.
The user inputs a query that attempts to execute multiple \visflow functions that exceed the limit expected by the grammar or the web client handler.
An example is \query{Highlight bachelors in red and masters in green in node-15}.
This is achievable using attribute filters to find candidates with Bachelors and Masters degrees, followed by visual editors to give them colors, and finally a set operator to merge the two groups.
In this case multiple \visflow functions have to be performed altogether.
The parser and execution handler did not expect queries of this composite level.
The grammatical structure between these multiple functions poses parsing difficulty.
It is recommended that composite queries are split into multiple smaller steps so as not to overload the NLI with a complicated grammatical structure that exceeds its parsing capability.

\item \myparagraph{Bug}.
The system should have the capability of handling that query.
But due to an implementation bug that was unidentified at the time of the user study,
the query parsing or execution went wrong and did not arrive at the expected outcome.

\end{itemize}